**Differential Multi-probe Thermal Transport Measurements of Multi-walled Carbon Nanotubes grown by Chemical Vapor Deposition**

Qianru Jia,[1] Yuanyuan Zhou,[1] Xun Li,[2] Lucas Lindsay,[2] Li Shi[1]*

[1]Walker Department of Mechanical Engineering, The University of Texas at Austin, Austin, Texas 78712, USA

[2]Materials Science and Technology Division, Oak Ridge National Laboratory, Oak Ridge, TN 37831, USA

*Email: lishi@mail.utexas.edu

*This manuscript has been authored by UT-Battelle, LLC under Contract No. DE-AC05-00OR22725 with the U.S. Department of Energy. The United States Government retains and the publisher, by accepting the article for publication, acknowledges that the United States Government retains a non-exclusive, paid-up, irrevocable, world-wide license to publish or reproduce the published form of this manuscript, or allow others to do so, for United States Government purposes. The Department of Energy will provide public access to these results of federally sponsored research in accordance with the DOE Public Access Plan (http://energy.gov/downloads/doe-public-access-plan).*

**Abstract**

Carbon nanotubes (CNTs) are quasi-one dimensional nanostructures that display both high thermal conductivity for potential thermal management applications and intriguing low-dimensional phonon transport phenomena. In comparison to the advances made in the theoretical calculation of the lattice thermal conductivity of CNTs, thermal transport measurements of CNTs have been limited by either the poor temperature sensitivity of Raman thermometry technique or the presence of contact thermal resistance errors in sensitive two-probe resistance thermometry measurements. Here we report advances in a multi-probe measurement of the intrinsic thermal conductivity of individual multi-walled CNT samples that are transferred from the growth substrate onto the measurement device. The sample-thermometer thermal interface resistance is directly measured by this multi-probe method and used to model the temperature distribution along the contacted sample segment. The detailed temperature profile helps to eliminate the contact thermal resistance error in the obtained thermal conductivity of the suspended sample segment. A differential electro-thermal bridge measurement method is established to enhance the signal-to-noise ratio and reduce the measurement uncertainty by over 40%. The obtained thermal resistances of multiple suspended segments of the same MWCNT samples increase nearly linearly with increasing length, revealing diffusive phonon transport as a result of phonon-defect scattering in these MWCNT samples. The measured thermal conductivity increases with temperature and reaches up to $390 \pm 20$ W m$^{-1}$ K$^{-1}$ at room temperature for a 9-walled MWCNT. Theoretical analysis of the measurement results suggests submicron phonon mean free paths due to extrinsic phonon scattering by extended defects such as grain boundaries. The obtained thermal conductivity is decreased by a factor of 3 upon electron beam damage and surface contamination of the CNT sample.

| Nomenclature | | | | |
|---|---|---|---|---|
| $A$ | cross-sectional area (m$^2$) | *Greek symbols* | | |
| $C$ | specific heat capacity (J m$^{-3}$ K$^{-1}$) | $\alpha$ | | chiral angle |
| $d$ | diameter (nm) | $\beta$ | | power law exponent |
| $G$ | thermal conductance (W K$^{-1}$) | $\hbar$ | | reduced Planck's constant |
| $g$ | mass variance parameter | $\lambda$ | | phonon coupling constant |

| | | | |
|---|---|---|---|
| $h$ | distance (nm) | $\tau$ | relaxation time (s) |
| $I$ | heating current (A) | $\omega$ | phonon angular frequency |
| $\kappa$ | thermal conductivity (W m$^{-1}$ K$^{-1}$) | $\theta, \phi$ | temperature rise (K) |
| $L$ | length (m) | *Subscripts* | |
| $n$ | phonon distribution function | $b$ | beam |
| $Q$ | heat flow rate (W) | $c$ | contact |
| $q$ | heat flux (W m$^{-2}$) | $d$ | defect |
| $R$ | thermal resistance (K W$^{-1}$) | $e$ | electrical |
| $T$ | temperature (K) | $h$ | high |
| $t$ | thickness (m) | $i$ | inner |
| $v$ | velocity (m s$^{-1}$) | $L$ | left |
| $w$ | width (m) | $l$ | low |
| | | $o$ | outer |
| | | $ph$ | phonon |
| | | $R$ | right |
| | | $s$ | support |
| | | $ref$ | reference |

# 1. Introduction

In nonmetallic solids, the thermal conductivity is dominated by the contribution from phonons, the energy quanta of lattice vibration waves. One-dimensional (1D) lattice with simplified interatomic potential can exhibit the Fermi-Pasta-Ulam-Tsingou (FPUT) paradox, for which recurrence of excited phonon modes is accompanied by extremely slow decay of heat flux correlation [1] and a divergence of the effective thermal conductivity ($\kappa$) with length ($L$) according to a power law relationship, $\sim L^{\beta}$ with $0 < \beta < 1$ [2–4] even at room temperatures. In three-dimensional bulk crystals, in comparison, the divergence is reduced or even removed by elastic anisotropy at low temperatures and by four-phonon scattering at high temperatures [5].

The solid thermal conductivity of graphitic materials is high and can potentially exhibit peculiar behaviors that deviate from Fourier's law description of diffusive thermal transport. Among different graphitic materials, carbon nanotubes (CNTs) are quasi-one dimensional nanostructures that can behave differently from idealized 1D lattice chains and 3D crystals. A first-principles calculation performed for the defect free single-walled CNT (SWCNT) demonstrates the importance of the inclusion of both Normal and Umklapp phonon-phonon scattering processes and obtains $\kappa$ values as high as 6000 W m$^{-1}$ K$^{-1}$ at room temperature [6]. According to several theoretical calculations [6–11], first-order three phonon scattering alone yields a $L^{1/2}$ divergence of the calculated $\kappa$ of a SWCNT, whereas additional higher-order phonon-phonon scattering results in a saturation of $\kappa$ when $L$ is over several microns.

In addition, two recent experiments have reported the observation of second sound, signature of hydrodynamic phonon transport, in graphite at temperatures between about 80 K and 120 K due to strong Normal scattering of flexural phonons in the layered materials [12,13].

Peculiar hydrodynamic phonon transport phenomena can potentially occur in CNTs at an intermediate temperature near 100 K according to a theoretical calculation [14].

These theoretical predictions have attracted interests in fundamental thermal transport investigations and thermal management applications of CNTs. However, current reported experimental thermal conductivity values of individual single-walled (SW) and multi-walled (MW) CNTs vary by one order of magnitude due to the variation of sample quality and limitations and uncertainties in the measurement methods [15–20]. Raman spectroscopy was employed to profile the temperature distribution along the electrically or optically heated suspended CNT samples and extract the contact thermal resistance and resolve the spatial variation of the thermal conductivity [21–24]. Micro-Raman measurements of long suspended CNTs have obtained high thermal conductivity values that saturates to about 2900 $Wm^{-1}$ $K^{-1}$ at room temperature as the length is increased to 10 μm [20]. Due to the limited temperature sensitivity of the Raman thermometry technique, the sample is often heated by the Raman laser beam by tens of degrees higher than the environment, making it difficult to use this method to measure the thermal conductivity at low temperatures to obtain the temperature-dependent behavior. In comparison, contact thermal resistance errors have been a major issue for past two-probe electro-thermal microbridge measurement of thermal transport in individual nanotubes. It was previously found that a large contact thermal resistance limited the measured two-probe thermal conductivity of a SWCNT and a double-walled CNT (DWCNT) sample suspended between two microfabricated resistance thermometers to about 600 W $m^{-1}$ $K^{-1}$ at room temperature [25].

Recently, a multi-probe thermal transport measurement has been reported as an effort to measure the contact thermal resistance and the thermal resistance of the suspended segment [26–29]. This method has been used to probe thermal transport of highly defective individual CNTs grown in the nanopores via catalyst-free chemical vapor deposition (CVD) method [30]. Besides minimizing the defects in the CNT samples, further advances in this and other experimental methods are required to enhance the measurement sensitivity for these nanostructures and completely eliminate the contact thermal resistance error in order to obtain accurate measurement data for comparison with theories of non-diffusive and quantized thermal transport behaviors of CNTs.

Here, we report advances in multi-probe thermal transport measurements of crystalline multi-walled CNTs grown by high-temperature CVD. The measured interfacial resistance is used in a detailed analytical model of the temperature profile along the contacted sample segment to completely eliminate the contact resistance error and obtain the true thermal conductivity of the suspended segment. A differential electro-thermal bridge measurement method is used to enhance the signal-to-noise ratio and reduce the measurement uncertainty by over 40%. The observed temperature-dependent thermal conductivity is analyzed by a theoretical model to extract submicron phonon scattering mean free paths, which are attributed to extrinsic scattering processes by extended defects such as grain boundaries in these CVD MWCNT samples. The enhanced measurement method is further used to reveal effects of electron damage and surface contamination on the thermal conductivity of suspended MWCNTs.

## 2. Experimental Methods

### 2.1 Carbon Nanotube Synthesis

**Figure 1(a)** shows 40-$\mu$m-wide trenches that were etched on a Si substrate with the use of Deep Reactive Ion Etching (RIE) [20]. Iron (Fe) catalyst with about 1 nm thickness was evaporated on one end of the Si substrate. Following a prior report [31], the as-prepared Si substrate was placed in a 0.3-inch-diameter quartz tube inside a 1-inch-diameter tube furnace to obtain laminar flow. The stabilized gas flow helps to lift the catalyst nanoparticles at the tip and grow long CNTs via a tip-growth process. After the catalysts were calcinated at 950 °C for 30 min with the $H_2$ and Ar gas mixture at the flowrates of 10 and 40 sccm, respectively, the $CH_4$, $H_2$, and Ar gases were flowed at 950 °C for 3 hrs at the rates of 5, 5, and 40 sccm at atmospheric pressure, respectively. The system was cooled down to the room temperature with the same gas flow. Horizontally aligned ultralong CNTs were grown across the trenches, with length up to about 100 µm. The diameter of the synthesized CNT is determined by the catalyst nanoparticle size [32].

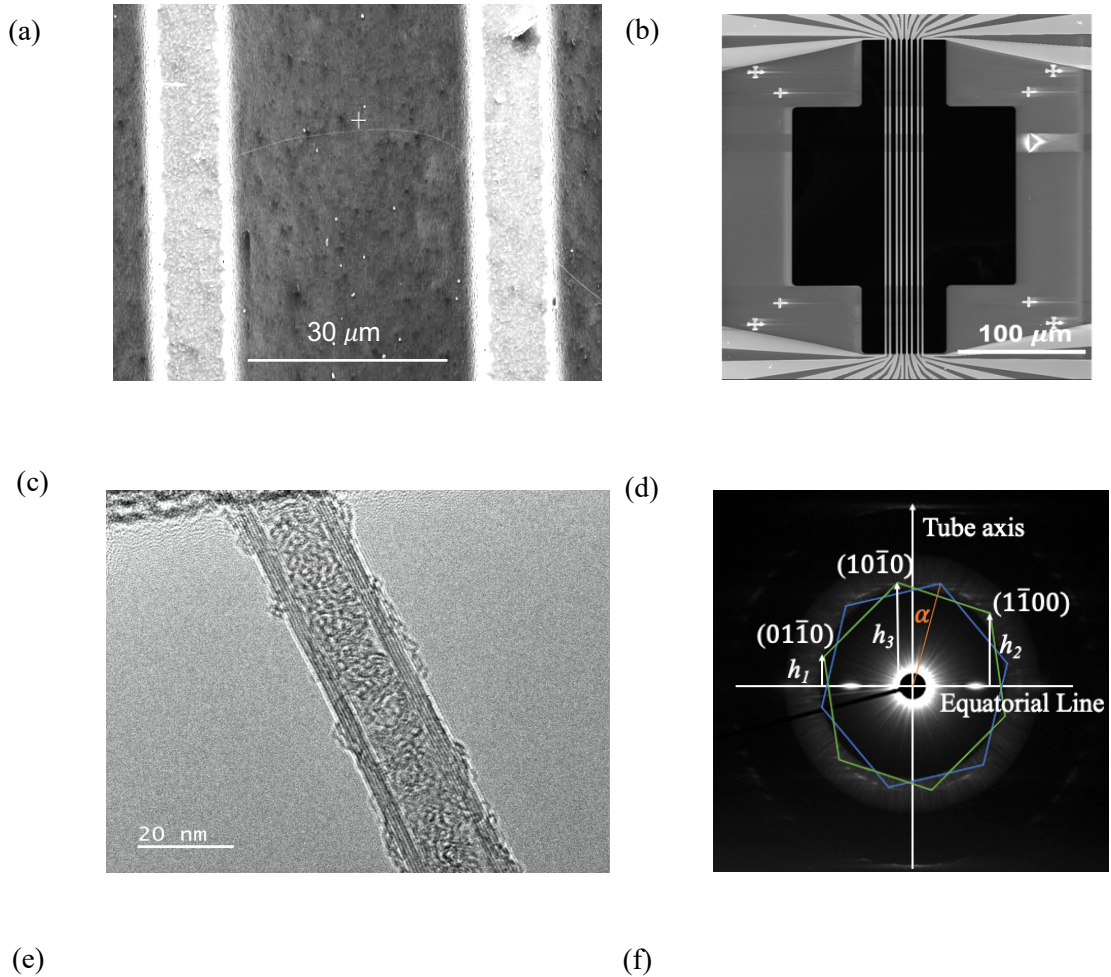

(a) (b)

(c) (d)

(e) (f)

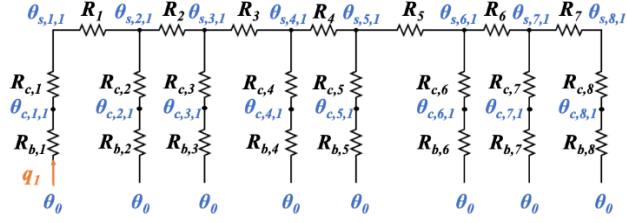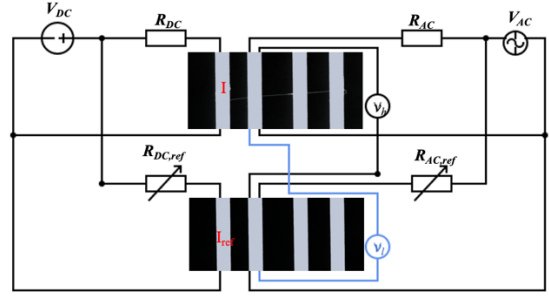

**Figure 1.** (a) Scanning electron microscopy (SEM) image of an isolated carbon nanotube (CNT) grown across an etched trench in a Si wafer. (b) SEM image of the measurement device that consists of eight 300-μm-long, 2-μm-wide suspended Pd/Cr/SiN$_x$ electrical heaters and resistance thermometers. (c) HRTEM image of MWCNT sample 1 assembled on the multiprobe device. (d) Selected area diffraction pattern for sample 1. The two hexagons represent the first order graphite-like $\{10\bar{1}0\}$ diffraction spots from the top and bottom of the same shell of CNT. Multiple additional spots are visible in the diffraction pattern and indicate that the sample contains multiple shells with different chiral angles. (e) Thermal resistance circuit of the measurement device when the first Pd/Cr/SiNx line is electrically heated at a Joule heating rate of $q_1$. $R_j$, $j = 1$ to 7, is the apparent thermal resistances of the $j^{th}$ suspended segments of the suspended sample. $R_{c,j}$, $j = 1$ to 8 is the contact thermal resistance between the $j^{th}$ Pd/Cr/SiN$_x$ line and the sample. $R_{b,j}$, $j = 1$ to 8 is the thermal resistance of the $j^{th}$ Pd/Cr/SiN$_x$ resistance thermometer line. $\theta_{c,j,i}$ and $\theta_{s,j,i}$ are respectively the Pd/Cr/SiN$_x$ line temperature rise and average sample temperature rise at the $j^{th}$ contact with the sample when the $i^{th}$ line is electrically heated. The temperature rise $\theta_0$ at the two ends of each of the suspended Pd/Cr/SiN$_x$ lines is assumed to be negligible. (f) Differential bridge measurement setup. A CNT is assembled only on the sample device (top panel) but not on the reference device (bottom panel). The heating to the two devices is adjusted by a variable resistor ($R_{DC,ref}$) so that the sensor temperature rise

$\theta_{j,i}$ on the reference device is the same as that of the sample device when no CNTs were assembled on either device. Another variable resistor ($R_{AC,ref}$) is adjusted to nullify the AC voltage drop difference across the two sensing lines ($V_s = V_h - V_l$) at zero DC heating voltage ($V_{DC} = 0$).

**2.2 Transmission Electron Microscopy (TEM) Measurements**

The measurement device in **Figure 1(b)** consists of eight suspended Pd/Cr/SiN$_x$ resistance thermometer (RT) lines. An isolated CNT sample was transferred from the growth substrate with deep trenches onto the microfabricated device with the use of a nanomanipulator. Upon the transfer, straight nanotube segments were suspended across the RT lines. Following the thermal transport measurement, high-resolution transmission electron microscopy (HRTEM) was used to characterize the structure of the suspended MWCNT sample assembled on the multiprobe device. According to the TEM measurements (**Figure 1(c)**), the outer diameter of these MWCNTs is in the range of 9.8-13.2 nm and the number of shells varies from 7 to 9, as summarized in **Table 1** for 3 samples. The selected area diffraction pattern (SAED) (**Figure 1(d)**) of sample 1 shows multiple equatorial lines that vary by a few degrees from the average orientation and streaks on spots, indicating the presence of a rotation angle between different layers. A chiral angle ($\alpha$) of about 9.2° can be calculated for one shell of sample 1 based on the relative distances of the peaks ($h_1$-$h_3$) from the equatorial line in SAED according to the following equation [33] and is independent of the scale or a diffraction astigmatism.

$$\alpha = \arctan\left(\frac{1}{\sqrt{3}} \frac{h_2 - h_1}{h_3}\right) = \arctan\left(\frac{1}{\sqrt{3}} \frac{2h_2 - h_3}{h_3}\right) \tag{1}$$

**Table 1** MWCNT sample dimensions measured by TEM. $d_i$ and $d_o$ are the inner and outer CNT diameters, respectively. $L_j$ is the suspended CNT length of the $j^{th}$ segment.

| Sample | # Shells | $d_i/d_o$ [nm] | $L_2$ [μm] | $L_3$ [μm] | $L_4$ [μm] | $L_5$ [μm] | $L_6$ [μm] |
|---|---|---|---|---|---|---|---|
| 1  | 7 | 8.5/13.2 | 8.2 | 5.0 | 6.3 | 2.2 | 1.0 |
| 10 | 9 | 3.7/9.8  | 3.0 | -   | -   | -   | -   |
| 11 | 9 | 3.7/9.8  | 3.0 | -   | -   | -   | -   |

### 2.3 Non-differential Multi-probe Thermal Transport Measurement

**Figure 1(e)** shows the thermal circuit of the measurement device. The measurement is performed with the sample in the vacuum space of the cryostat. When a single RT line $i$ is electrically heated at a rate of $q_i$, the electrical resistances of each RT line $j$ is measured to obtain its temperature rise ($\theta_{j,i}$) based on its temperature coefficient of resistance. The measurement is repeated for each RT to be used as the heater line to obtain a set of total n x n data on $\theta_{j,i}/q_i$, where both $i$ and $j$ run between n=1 and n = 8 for the eight-probe sample shown in **Figure 1(b, e)**. This data set can be used to obtain the thermal resistance ($R_{b,j}$) of each RT, the thermal resistance $R_j$ ($j$=2 to 6) and contact thermal resistance $R_{c,j}$ ($j$ =2 to 7) of each inner suspended segment of the nanostructure [30]. Here, the thermal resistance $R_j$ can be separated from the contact thermal resistances because the heat flow across them is different. In comparison, the thermal resistance of the two end segments $R_j$ ( $j$ = 1 or 8) cannot be identified because the heat flow through the outer segment is the same as that through the corresponding end contact ($R_{c,j}$, $j$ = 1 or 8) [26–29]. In addition, the background signal caused by parasitic heat transfer between the heating line and other RTs due to radiation, residual gas molecules inside the evacuated cryostat stage, and non-zero thermal resistance of the silicon substrate was measured prior to the transfer of the nanotube sample on the multi-probe device. The measured background is then subtracted from the $\theta_{j,i}/q_i$ signal measured on the device with the nanostructure sample to obtain the thermal conductance in the suspended segment of the sample. As in prior works [30], the Joule heating current in the $i^{th}$ line is discretely stepped according to a sinusoidal function. Fast Fourier transform (FFT) data analysis is applied to find the second harmonic component by analyzing the measured stepped DC heating current and voltage drop in the heating line in addition to the AC voltage drop along other thermometer lines with a small AC sensing current. The FFT results of the sample and background measurement are shown in **Figure 2(a)**.

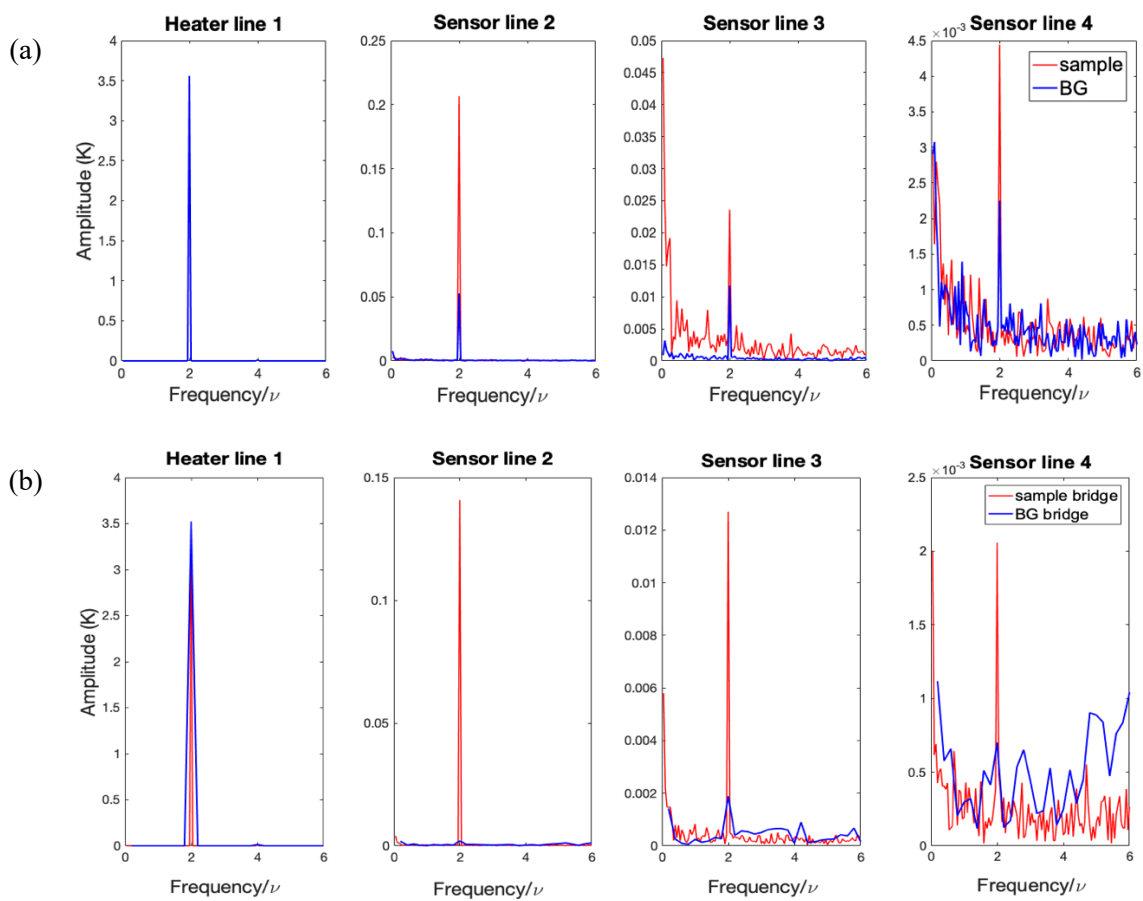

**Figure 2**. Fast Fourier Transform analysis of the measured thermometer temperature modulation from (a) separate non-differential measurements of the sample device (red) and background device (blue), and from (b) differential bridge measurements of a sample-blank bridge (red) and the background blank-blank bridge (blue). Amplitude spectrum of the measured average temperature modulation of each thermometer line is labeled by the thermometer line number in each panel when line 1 is electrically heated at 16 different constant current levels that were cycled to follow a sinusoid at frequency ν < 0.025 Hz for CNT sample 11. The sample stage temperature is 300 K.

**2.4 Differential Multi-probe Thermal Transport Measurements**

Instead of measuring the background signal separately and then subtracting it from the nanotube measurement data to obtain the contribution from the nanotube, in this work we establish a differential multi-probe thermal transport measurement method to eliminate the background thermal conductance directly during the measurement and to enhance the signal to noise ratio. In a prior work [34], a similar differential bridge measurement can achieve a temperature sensitivity of $10^{-4}$ K by greatly reducing the common-mode noise including substrate temperature fluctuation. The differential measurement obtains the relative increase in the sensing RT temperature for the sample device compared to that for a blank reference device. This increase is caused solely by heat conduction through the nanotube sample. The schematic of the Wheatstone bridge circuit is shown in **Figure 1(f)**. The blank reference device shares the same design as and is adjacent to the sample device on the same chip, except that there is no CNT sample assembled on the reference device. During the differential measurement, the two variable resistors, $R_{DC}$ and $R_{DC,ref}$, are adjusted to achieve the following ratio between the heating currents ($I$ and $I_{ref}$) in the sample device and the reference device

$$\frac{I}{I_{ref}} = [\frac{R_{b,i,ref}R_{b,j,ref}}{R_{b,i}R_{b,j}}\frac{R_{e,i,ref}}{R_{e,i}}]^{\frac{1}{2}} \quad (2)$$

where $R_{e,i}$ is the electrical resistance of the $i^{th}$ heater line, the *ref* subscript is used to differentiate the reference blank device from the sample device, and the beam thermal resistances $R_{b,j}$ are obtained from the non-differential thermal transport measurement for both the sample device and the reference device. Under this heating condition [34], the same temperature rise ($\theta_{s,j,i}$) would be obtained on the $j^{th}$ RT line of both the sample device and the reference device when the nanotube sample was absent on both devices.

In addition, two lock-in amplifiers are used to measure the difference in the AC voltage drops along the $j^{th}$ RT lines of the sample and reference devices as $V_s = V_h - V_l$, as shown in **Figure 1(f)**. A variable resistor $R_{AC,ref}$ is used to nullify $V_s$ when there is no DC heating in the two devices. With this balance, the variation of $V_s$ during heating is caused by the different temperature rise on the $j^{th}$ sensing line between the sample and reference devices as a result of heat conduction in the CNT sample. When this differential measurement method is used to test two blank devices without a nanotube sample, the obtained background blank-blank bridge measurement results are compared in **Figure 2(b)** with the sample-blank bridge measurement where the nanotube is assembled on the sample device. The vanishing signal measured on the blank-blank bridge device shows the effectiveness of this method for eliminating the parasitic background and provides an evaluation of the systematic uncertainty of this measurement method, as shown in **Table 2**. Besides being able to correct for the background signal like the nondifferential measurement, the differential bridge measurement improves the signal-to-noise ratio in the measurement of the small sensor temperature rise caused by heat conduction through the nanotube sample.

**Table 2.** FFT amplitude of the temperature rises of four-probe measurement devices in both non-differential measurements of a sample and a blank device (Sample-blank), differential measurements between one sample device and one blank device (Sample-blank bridge), and differential measurements between two blank devices (Blank-blank bridge). The stage temperature is at 300K.

|  | Heater 1 | Sensor 2 | Sensor 3 | Sensor 4 |
|---|---|---|---|---|
| Sample (K) | 3.52 | 0.21 | 0.026 | 0.0044 |
| Blank (K) |  | 0.05 | 0.012 | 0.0023 |
| Sample-blank (K) |  | 0.16 | 0.014 | 0.0021 |
| Sample-blank bridge (K) | 3.52 | 0.14 | 0.013 | 0.0021 |
| Blank-blank bridge (K) | 3.52 | 0.00042 | 0.00070 | 0.00021 |
| Systematic error |  | 0.30% | 8.9% | 10.3% |

## 3. Contact Temperature Profile Model

The resistance thermometry measurements obtain the average temperature rise $\theta_{j,i}$ along each RT line for a given Joule heating rate $q_i$ in the $i^{th}$ heater line. As shown in prior reports [26,30], the measurement data can be used to extract the average contact temperature rises of both the RT line and the nanotube sample, $\theta_{c,j}$ for $j=1$ to $n$ and $\theta_{s,j}$ for $j=2$ to $n-1$ where we have eliminated the subscript $i$ used in $\theta_{j,i}$ to represent the heater line. Due to the finite contact length, these average sample temperature rises ($\theta_{s,j}$ and $\theta_{s,j+1}$) at the two adjacent contacts differ from those ($\theta_{L,j}$ and $\theta_{R,j}$) at the left and right ends of the $j^{th}$ supported sample segment. Due to this difference, the measured apparent thermal resistance of an inner suspended sample segment, $R_j = (\theta_{s,j} - \theta_{s,j+1})/Q_j$, still deviates from the actual thermal resistance of the suspended segment, $R'_j \equiv (\theta_{R,j} - \theta_{L,j+1})/Q_j$, where $Q_j$ is the heat flow in the $j^{th}$ suspended sample segment. In addition, due to the different heat current ($Q_{c,j}$) through the contact between the $j^{th}$ suspended Pd/Cr/SiN$_x$ beam and the suspended CNT sample compared to $Q_j$ in the adjacent suspended sample segment, $R'_j \neq R_j - R_{c,j} - R_{c,j+1}$.

In this work, we use the measured sample-support interface thermal resistance $R_{c,j}$ to model the temperature profile along the supported sample segment and evaluate the $R'_j/R_j$ ratio. For the supported segment of the nanotube, we express the phonon Boltzmann transport equation (BTE) as

$$\boldsymbol{v} \cdot \nabla n = \frac{n_0(T) - n(\mathbf{k},\mathbf{r})}{\tau} + \frac{n_0(T_c) - n(\mathbf{k},\mathbf{r})}{\tau_s} \tag{3}$$

where $\boldsymbol{v}$ is the phonon group velocity. In this expression, the non-equilibrium distribution ($n$) of phonons in the real ($\mathbf{r}$) and momentum ($\mathbf{k}$) space relaxes to a local equilibrium distribution ($n_0$) characterized with a temperature $T$ via phonon scattering within the nanotube over a time scale of $\tau$. In addition, scattering between phonons across the sample-support interface relaxes $n(\mathbf{k},\mathbf{r})$ toward $n_0(T_c)$ at the contact temperature ($T_c$) of the support. We approximate $\nabla n \approx \frac{\partial n_0}{\partial T} \nabla T$ to obtain

$$n(\mathbf{k},\mathbf{r}) = \frac{n_0(T)}{1+\tau/\tau_s} + \frac{n_0(T_c)}{1+\tau_s/\tau} - (\tau^{-1} + \tau_s^{-1})^{-1} \frac{\partial n_0}{\partial T} \boldsymbol{v} \cdot \nabla T \tag{4}$$

With $n_0$ given by the Bose-Einstein distribution that does not produce a net heat flux, the heat flux along the nanotube is obtained as

$$\boldsymbol{q} = \frac{1}{V}\sum_\mathbf{k} \hbar\omega(\mathbf{k}) n(\mathbf{k},\mathbf{r}) \boldsymbol{v} = -\frac{1}{V}\sum_\mathbf{k} \hbar\omega(\mathbf{k})(\tau^{-1} + \tau_s^{-1})^{-1} \frac{\partial n_0}{\partial T}(\boldsymbol{v} \cdot \nabla T)\boldsymbol{v} \tag{5}$$

where $V$ is the nanotube volume, $\hbar$ is the reduced Planck's constant, and $\omega$ is the phonon angular frequency. The thermal conductivity of the supported segment of the nanotube is

$$\kappa_s \equiv -\frac{\boldsymbol{q}}{\nabla T} = \frac{1}{V}\sum_\mathbf{k} \hbar\omega(\mathbf{k})(\tau^{-1} + \tau_s^{-1})^{-1} \frac{\partial n_0}{\partial T} \boldsymbol{v} \cdot \boldsymbol{v} \tag{6}$$

The $\tau_s$ term in this equation can be removed to obtain the thermal conductivity ($\kappa$) of the suspended sample segment [35].

In addition, we perform the following summation of the BTE over the different **k** states in the first Brillouin zone

$$\sum_{\mathbf{k}} \hbar\omega(\mathbf{k})\mathbf{v} \cdot \nabla n = \sum_{\mathbf{k}} \hbar\omega(\mathbf{k}) \frac{n_0(T) - n(\mathbf{k},\mathbf{r})}{\tau} + \sum_{\mathbf{k}} \hbar\omega(\mathbf{k}) \frac{n_0(T_c) - n(\mathbf{k},\mathbf{r})}{\tau_s} \quad (7)$$

The first term on the right-hand side of this equation vanishes because phonon scattering within the nanotube conserves the energy $U = \sum_{\mathbf{k}} \hbar\omega(\mathbf{k})n(\mathbf{k},\mathbf{r}) = \sum_{\mathbf{k}} \hbar\omega(\mathbf{k})n_0(T)$. When the frequency dependence of $\tau_s$ is ignored and the volumetric specific heat is defined as $C = \frac{dU}{VdT}$, this equation is reduced to

$$\nabla \cdot \mathbf{q} = \frac{C}{\tau_s}(T_c - T) \quad (8)$$

The right-hand side of this equation describes interfacial heat transfer characterized with a sample-support phonon coupling constant $\lambda_s \equiv C/\tau_s$. This equation derived from the BTE is essentially the same as the following heat diffusion equation in a fin

$$A\nabla \cdot \mathbf{q} = \frac{1}{R_{c,j} w_j}(T_c - T) \quad (9)$$

where $A$ is the cross-section area of the nanotube, $w_j$ is the contact length between the nanotube and the $j^{th}$ support, and $R_{c,j} = (\lambda_s A w_j)^{-1}$ is the contact thermal resistance between the nanotube and the support. The variation of the temperature rise ($\theta$) along the sample segment in contact with the $j^{th}$ thermometer is thus governed by

$$\kappa_s \frac{d^2\theta}{dx^2} = \lambda_s(\theta - \theta_{c,j}) \text{ for } -\frac{w_j}{2} \leq x \leq \frac{w_j}{2} \quad (10)$$

The solution takes the following form

$$\phi = \frac{\phi_{L,j} + \phi_{R,j}}{2\cosh\frac{w_j}{2L_s}} \cosh\frac{x}{L_s} + \frac{\phi_{R,j} - \phi_{L,j}}{2\sinh\frac{w_j}{2L_s}} \sinh\frac{x}{L_s} \quad (11)$$

where $\phi \equiv \theta - \theta_{c,j}$, $\phi_{L,j} \equiv \theta_{L,j} - \theta_{c,j}$, $\phi_{R,j} \equiv \theta_{R,j} - \theta_{c,j}$, and $L_s = (\kappa_s/\lambda_s)^{1/2}$ is the sample-support thermalization length. The average temperature rise of the sample segment in contact with the $j^{th}$ thermometer line can be obtained from an integration of Equation 11 as

$$\theta_{s,j} = \theta_{c,j} + \frac{L_s}{w_j}(\phi_{L,j} + \phi_{R,j})\tanh\frac{w_j}{2L_s} \tag{12}$$

We set up the following $2n$ boundary equations to describe continuous heat flow across each junction between the supported sample segment and suspended sample segment

$$-\kappa_s A \frac{d\phi}{dx} = Q_{j,L \text{ or } R} \tag{13}$$

As the left ($L$) end and right ($R$) end of the whole sample are adiabatic,

$$Q_{1,L} = Q_{n,R} = 0 \tag{14}$$

At the junction between a supported sample segment and an inner suspended sample segment,

$$Q_{j,R} = Q_{j+1,L} = \frac{\theta_{s,j} - \theta_{s,j+1}}{R_j} \text{ for } j=2 \text{ to } n\text{-}2 \tag{15}$$

In comparison, at the junction between the first supported sample segment and the 1st suspended segment,

$$Q_{1,R} = \frac{\theta_{c,1} - \theta_{s,2}}{R_1 + R_{c,1}} \tag{16}$$

Similarly, at the junction between the $n^{th}$ supported segment and the $(n\text{-}1)^{th}$ suspended segment,

$$Q_{n,L} = \frac{\theta_{s,n-1} - \theta_{c,n}}{R_{n-1} + R_{c,n}} \tag{17}$$

This set of $2n$ boundary conditions can be arranged into linear equations that can be used to solve for the $2n$ values of $\phi_{L,j}$ and $\phi_{R,j}$. The coefficients in this equation contain the measured values of $(R_1 + R_{c,1})$, $(R_{n-1} + R_{c,n})$, $R_j$ for $j=2$ to $n$-2, and $\theta_{c,j}$ for $j=1$ to $n$ [26,30], as well as the unknown $\kappa_s$.

Based on the measured $R_{c,j}$, we obtain $\tau_s = C/\lambda_s$, where the lattice specific heat $C$ is available from lattice dynamical calculation. In the diffusive phonon transport regime, $\kappa = R'_j L_j / A$, with $L_j$ being the length of the suspended sample segment. The temperature-dependent $\kappa$ can be used to extract the frequency dependent $\tau$, which can then be used together with $\tau_s$ to evaluate $\kappa_s$. Because the measurement can obtain $R_j$ instead of $R'_j$ directly, an iteration procedure is used to obtain a new $R'_j$ based on the measured $R_j$ and the $R'_j/R_j$ ratio from the previous iteration step. The new $R'_j$ is then used to evaluate $\kappa$, frequency-dependent $\tau$, and $\kappa_s$. The obtained $\kappa_s$ is subsequently used with the other measured values to obtain the values of $\phi_{L,j}$ and $\phi_{R,j}$, which can be used to calculate a new $R'_j/R_j = (\theta_{R,j} - \theta_{L,j+1})/(\theta_{s,j} - \theta_{s,j+1})$. The iteration is completed when the calculated $R'_j/R_j$ converges. **Figure 3(a)** shows the calculated temperature distribution along a MWCNT sample suspended on four thermometer lines based on this procedure and the measurement results. The inset shows how the nanotube temperature is thermalized toward the contact temperature for the second contact. Due to the large $R_j$ compared to $R_b$, $\theta_{c,4}/\theta_{c,1}$ is as small as 0.0021% when the 1st RT line is used as the heater line. Accurate measurement of the small temperature signals requires the use of the sensitive differential bridge measurement method.

(a)

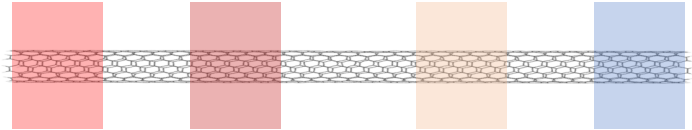

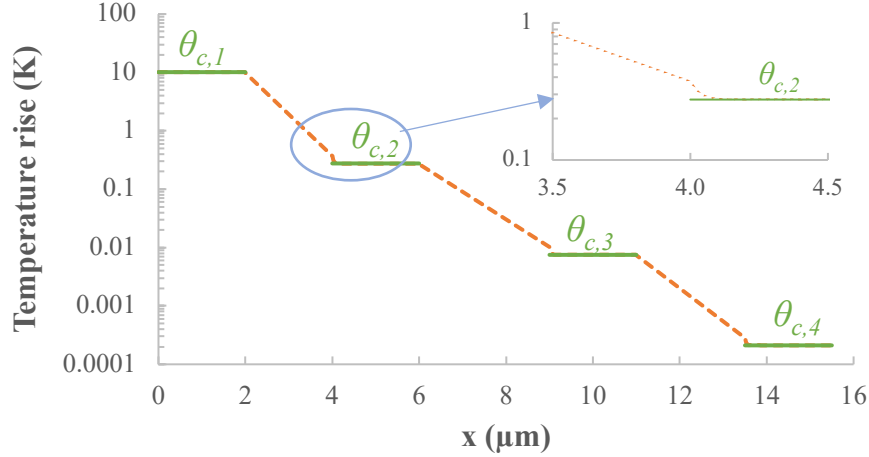

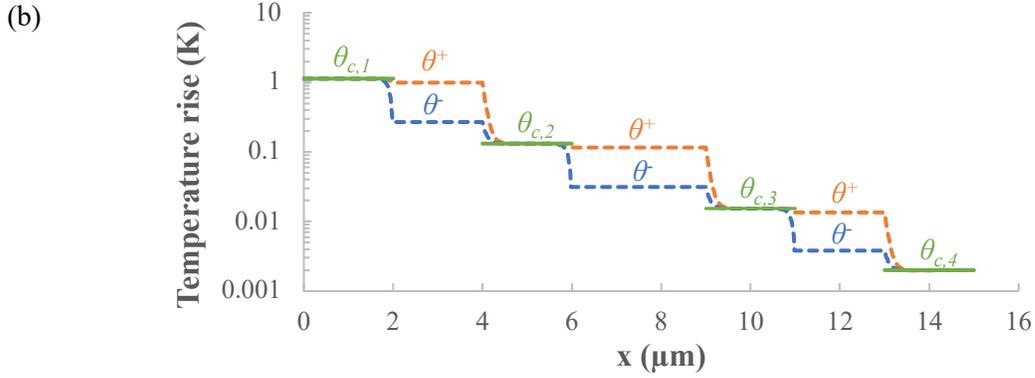

**Figure 3**. Calculated temperature profiles along (a) a measured diffusive MWCNT sample and a (b) hypothetical ballistic sample suspended on four RT lines.

We note that this model can be further extended to account for ballistic phonon transport in the suspended sample segment, where the right-moving phonons and the left-moving phonons take different temperatures ($\theta^+$ and $\theta^-$) that are spatially invariant along each suspended sample segment. In this case, Equation 10 and 11 are expressed separately for $\theta^+$ and $\theta^-$ to contain 4n values of $\theta^+_{L,j}, \theta^-_{L,j}, \theta^+_{R,j}, \theta^-_{R,j}$. The thermal resistance of the suspended ballistic segment is $R'_j = 1/G_b$, where $G_b$ is the ballistic thermal conductance. In addition, $R'_j/R_j = (\theta^+_{R,j} - \theta^-_{L,j+1})/(\theta_{s,j} - \theta_{s,j+1})$. These relations can be used to modify the 2n boundary equations for the diffusive phonon transport case. At both the leftmost and rightmost ends of the sample, four boundary conditions can be used to describe vanishing gradients in both $\theta^+$ and $\theta^-$, instead of just the two boundary equation 14 for the diffusive case. Moreover, additional 2(n-1) boundary equations can be expressed as

$$\theta^+_{R,j} = \theta^+_{L,j+1} \text{ and } \theta^-_{R,j} = \theta^-_{L,j+1} \text{ for } j = 1 \text{ to } n\text{-}1 \tag{18}$$

**Figure 3(b)** shows the calculated temperature distribution along a hypothetically ballistic nanotube sample suspended over four thermometer lines.

This model can also handle both strong and weak support scattering situations for both the diffusive and ballistic cases. In the limit of vanishing support scattering of phonons in the nanotube, $g_s$ approaches zero. In this case, Equation 11 is reduced to a linear temperature profile along the supported sample segment for the diffusive case and spatially invariant $\theta^+$ and $\theta^-$ for the ballistic case.

## 4. Results and Discussion

Several MWCNT samples have been used to establish this measurement method, as listed in **Table 1.** For sample 1 that is measured with the use of an 8-probe device and the non-differential measurement method, the measured $R_j$ increases with the suspended segment length of the MWCNT sample, as shown in **Figure 4**. The deviation of the measurement data from the linear fitting line is generally small and can be attributed to a variation of the crystal quality along different suspended segments. In addition, the resistance of the Pd/Cr/SiN$_x$ resistance thermometer starts to show nonlinear dependence on the temperature as the temperature is reduced to 50 K, which is higher than those found for thin film Pd resistance thermometers used in some prior experiments [36]. For this reason, the measurement data reported here is in a temperature range above 50 K where a linear temperature dependence of the Pd/Cr/SiN$_x$ resistance is observed.

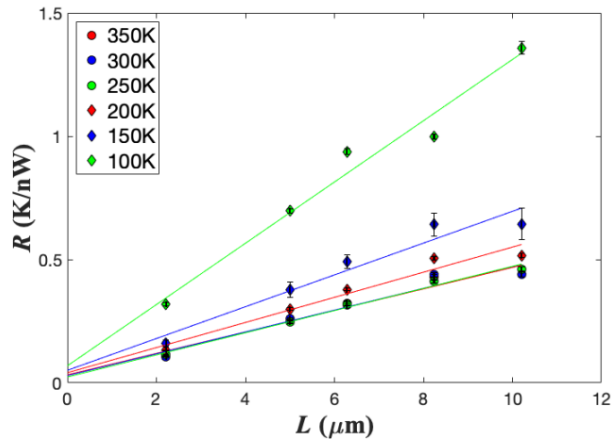

**Figure 4**. Measured thermal resistances ($R_j$) of the suspended segments of MWCNT sample 1 as a function of the suspended segment length at different temperatures specified in the legend.

Among these samples, sample 10 and sample 11 share the same four-probe measurement device design and were both measured via differential and non-differential multi-probe methods. The comparison of the measured thermal resistances of the MWCNT samples as a function of temperature of sample 10 and 11 are shown in **Figure 5 (a, b)**. The results from the differential and non-differential measurement methods are similar. They both show that the contact thermal resistances ($R_{c2}$ and $R_{c3}$) are more than one-order-of-magnitude smaller than the thermal resistance ($R_2$) of the middle suspended segment. Based on **Figure 5**, the measurement uncertainty is reduced by the differential method compared to the non-differential measurement method. For sample 11 at 250 K, the differential measurement method obtains the thermal resistance as $(0.422 \pm 0.005) \times 10^9$ K W$^{-1}$, which shows nearly 75% reduction in the uncertainty compared to the $(0.42 \pm 0.02) \times 10^9$ K W$^{-1}$ value obtained from the non-differential measurement. The minimum uncertainty reduction of 40% is found for sample 10 at 250 K, where the differential and non-differential measurement methods obtain the thermal resistance as (0.155 ± 0.004)x10$^9$ K W$^{-1}$ and (0.168 ± 0.007)x10$^9$ K W$^{-1}$, respectively. The reduced uncertainty is attributed to the elimination of some common mode noises, such as that caused by the fluctuation in the sample stage temperature, by the differential measurement method.

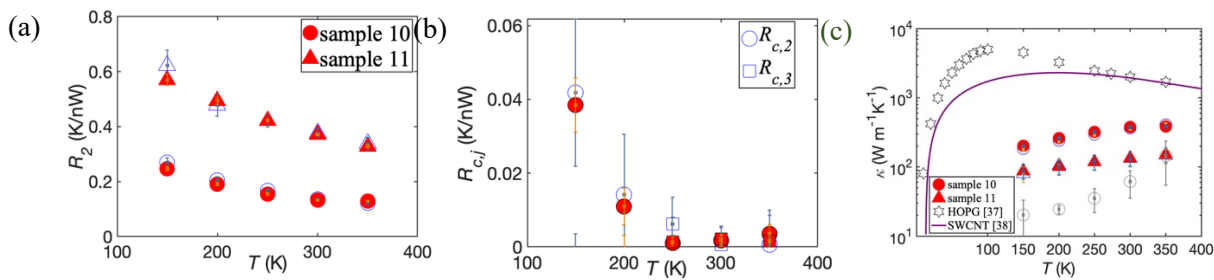

**Figure 5. (a)** Measured thermal resistances ($R_2$) of the middle-suspended segments of sample 10 and 11as a function of temperature. (b) Measured contact thermal resistances at the two inner thermometer lines of sample 11. (c) Measured thermal conductivity of the middle-suspended CNT segments of sample 10 and 11 as a function of temperature. Also shown for comparison are the highest reported graphite thermal conductivity values (star) included in reference [37] and the theoretical thermal conductivity calculated in reference [38] for a 3-μm long, (10,10) SWCNT with naturally occurring isotope concentrations. In all three panels, the data obtained with the differential method and the non-differential method are shown as filled and unfilled color markers, respectively. The gray circles show the results obtained after metal evaporation at the contacts of sample 10.

Based on these measurement results and the temperature profile model, we obtain that $R'_2/R_2$ is close to 97.2% and 99.0% for sample 10 and sample 11, respectively. For converting the obtained $R_2'$ to the thermal conductivity, we calculate the cross section area of the nanotube according to [25]

$$A \equiv \sum_j^m \pi d_j \delta = m\pi\delta[d_1 + (m-1)\delta] \tag{19}$$

where $m$ is the number of shells of the nanotube, $d_j$ is the diameter of the $j^{th}$ nanotube shell, and $\delta = 3.35 \text{Å}$ is the interplanar spacing of graphite. We note that this interplanar spacing of graphite is very close to the calculated result based on the outer ($d_o$) and inner ($d_i$) diameter measurement $(d_o - d_i)/2m$. The difference between this expression from another definition $A = \pi^2(d_o^2 - d_i^2)/4$ is within the reported measurement uncertainty.

As shown in **Figure 5(c)**, the obtained thermal conductivity of the suspended middle segment increases with increased temperature in the range between 100 and 350 K. The thermal conductivity of sample 10 reaches up to $390 \pm 20$ W m$^{-1}$ K$^{-1}$ at 300K. Lower thermal conductivity values are found in the other samples. In comparison, the thermal conductivity of high quality highly oriented pyrolytic graphite (HOPG) sample reaches 2000 W m$^{-1}$ K$^{-1}$ at room temperature, and peaks at a temperature near 100 K [37]. Similarly, the calculated theoretical thermal conductivity of a 3-μm long (10,10) SWCNT with naturally occurring isotope concentrations is about 1800 W m$^{-1}$ K$^{-1}$ at room temperature and shows a peak around 150 K [38]. The suppressed magnitude and upshift of the peak temperature observed in these MWCNT samples are attributed to extrinsic phonon scattering by defects that dominate intrinsic phonon-phonon scattering in these MWCNT samples. Due to the dominance of these extrinsic scattering mechanisms in the suspended segment of the nanotube, the additional support scattering in the supported segment produces a negligible effect on $\kappa_s$, which can be assumed as the same as $\kappa$ in the calculation of $R_2'/R_2$.

For sample 10, the thermal transport measurement was repeated after the sample was characterized with TEM at an electron beam energy of 120 keV and subsequent evaporation of 30 nm thick Pd/Cr through the open windows of a suspended $SiN_x$ mask onto the four contact areas between the MWCNT sample and the four thermometer lines. Based on the two TEM images in **Figure 6**, these additional experiments result in apparent amorphous coating on the MWCNT surface. This amorphous coating is due to either a spreading of the evaporated Pd/Cr onto the suspended nanotube segment or surface contamination during the additional TEM step. This structure change is accompanied with a large reduction of the measured thermal conductivity, as shown by the gray symbols of **Figure 5(c)**. In a prior work [25], the measured two-probe thermal conductivity of MWCNT was found to be reduced upon TEM measurement of the sample, and the reduction was attributed to electron beam irradiation damage of the sample. Besides the electron beam damage, the amorphous coating on the nanotube surface can increase the mass and scatter phonons in the nanotube to reduce its thermal conductivity, similar to scattering of graphene phonons by the substrate in supported graphene [39] and deposited metal coating on suspended graphene [40]. Because the amorphous coating covers a large faction of the nanotube surface, they can produce a more pronounced effect on phonon transport than the interaction between the supported nanotube and the thermometer through a narrow contact width. Both the amorphous coating and the electron beam induced damage can be responsible for the large reduction of the thermal conductivity.

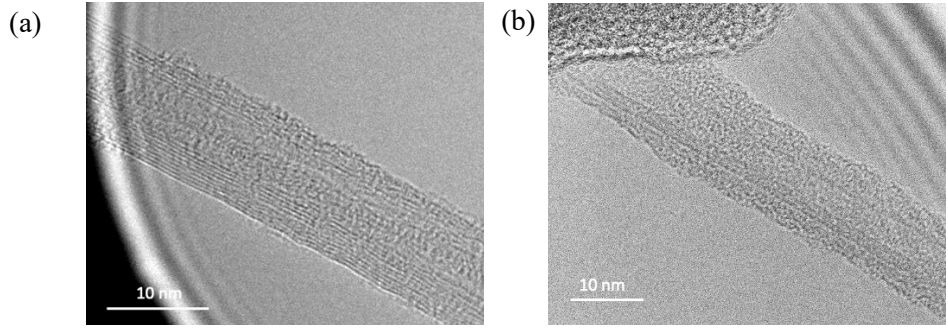

**Figure 6.** TEM images of sample 10 before (a) and after (b) metal evaporation through a shadow mask onto the four contact areas between the MWCNT and the thermometer lines.

## 5. Theoretical Analysis

We have carried out first-principles-based theoretical calculations to better understand the extrinsic phonon scattering mechanism behind the observed thermal conductivity of the MWCNT samples. As the inner diameters of samples 10 and 11 are greater than 3.5 nm we do not expect that nanotube curvature plays a significant role in determining their thermal conductivities [41]. In addition, as each sample is 9 layers thick we expect that transport along the nanotube axis behaves like that of bulk graphite [42]. With these considerations models of defect-derived thermal conductivity, $\kappa = \frac{1}{V}\sum_{\mathbf{k}} C(\mathbf{k})v^2(\mathbf{k})\tau(\mathbf{k})$, were built using harmonic and anharmonic phonon properties of bulk graphite calculated in a recent study [13]. A uniform $q$ mesh of $19 \times 19 \times 7$ is used with an adaptive Gaussian broadening scheme that is implemented in the ShengBTE package [43]. Here, $C(\mathbf{k})$, $v(\mathbf{k})$, and $\tau(\mathbf{k})$ are mode-specific specific heat, velocity component along the temperature gradient, and relaxation time for phonon mode $\mathbf{k}$.

To examine the effect of diffuse boundary scattering from polymer residue and other defects at the exterior surface, we solved the steady-state Peierls-Boltzmann equation for finite size samples with a deviational Monte Carlo method. Details of the method can be found in prior reports [13,44–47]. **Figure 7(a)** shows the thermal conductivity for 9-layer thick and infinitely wide samples with different lengths along the transport direction. Diffuse boundary scattering at the top surface considerably reduces the thermal conductivity, but it alone cannot explain the temperature-dependent measurements of 3 *μm* long samples. In comparison, the measured data is reproduced when the simulated sample lengths are reduced to about 400 and 60 nm for samples 10 and 11, respectively. In experiments, an effective sample length reduction may be caused by the existence of extended defects such as grain boundaries associated with dislocations, which have been observed in TEM measurements of MWCNT samples grown with a similar CVD recipe due to dynamic reshaping of the catalyst particle during the growth process [25]. Thick MWCNT samples grown with the thermal CVD method tend to be more defective than those produced by laser ablation [48] and arc-discharge processes [49,50].

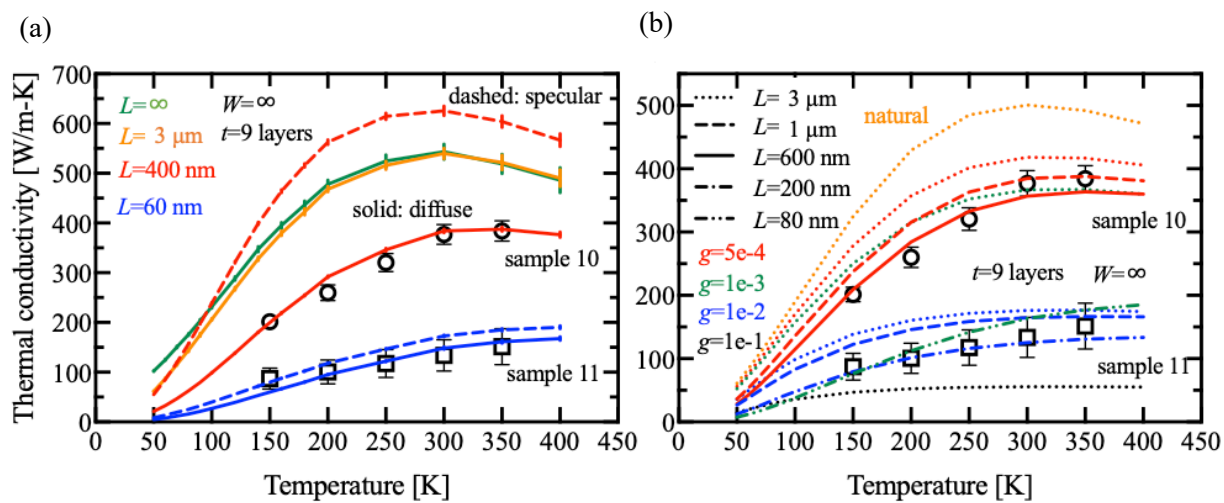

**Figure 7.** (a) Density Functional Theory (DFT)-derived Monte Carlo simulation of thermal conductivity for 9-layer thick and infinitely wide graphite samples with different lengths indicated in the legend. Solid curves represent simulations with diffuse boundary scattering on the top surface and specular reflection on the bottom surface. Dashed curves represent specular reflection

on both surfaces. (b) Calculated thermal conductivity with intrinsic phonon scattering, point-defect scattering, and empirical boundary scattering for different values of $g$ and $L$.

We also examined the effect of phonon-point-defect scattering from first principles in conjunction with empirical boundary scattering. Here, phonon lifetimes, $\tau(\mathbf{k})$, are calculated using Matthiessen's rule as

$$\tau^{-1}(\mathbf{k}) = \tau_{ph}^{-1}(\mathbf{k}) + \tau_d^{-1}(\mathbf{k}) + \tau_{boundary}^{-1}(\mathbf{k}) \tag{20}$$

where subscripts *ph*, *d*, and *boundary* represent intrinsic three-phonon, phonon-point-defect, and phonon-boundary scatterings, respectively. The phonon-point-defect scattering rate is modeled via a mass-variance scattering model

$$\tau_d^{-1}(\mathbf{k}) = \frac{\pi \omega^2(\mathbf{k})}{2} \sum_i g(i) |\mathbf{e}_\mathbf{k}^*(i) \mathbf{e}_{\mathbf{k}'}(i)|^2 \delta[\omega(\mathbf{k}) - \omega(\mathbf{k}')] \tag{21}$$

where $\mathbf{e}_\mathbf{k}(i)$ is the eigenvector for atom $i$, and $g(i)$ is the mass variance parameter [51,52]. The boundary scattering rate is $\tau_{boundary}^{-1}(\mathbf{k}) = L^{-1} v(\mathbf{k}) + t^{-1} v_z(\mathbf{k})$ where $t = 6$ nm representing a 9-layer sample with diffuse scattering on only one of the two surfaces, and $v_z(\mathbf{k})$ is the mode group velocity in the thickness direction. We tuned $g(i)$ beyond that of naturally-occurring isotope abundances and $L$ in this calculation to fit the measured data as shown in **Figure 7(b)**. The dotted curves suggest that point defect scattering alone cannot lead to the measured temperature dependence of thermal conductivity. However, together with reduced grain boundary scattering lengths $L$, the measured thermal conductivities are well matched.

## 6. Conclusion

This work makes several advances in multi-probe measurement of thermal transport properties of CNTs. The differential Wheatstone bridge measurement is able to enhance the signal to noise ratio and to reduce the uncertainty to less than 60% of the non-differential multi-probe measurement. The measured sample-support interfacial resistance is used in an analytical model of the temperature profile along the supported sample segment to fully eliminate the effect of contact temperature drop and obtain the true intrinsic thermal conductivity of the suspended segment. As the contact thermal resistance error is eliminated, the measured thermal conductivity reveals a sensitive dependence on electron beam damage and surface contamination of the nanotube sample. Based on the theoretical analysis, the thermal conductivity of these CVD MWCNT samples is limited by extrinsic phonon scattering by extended defects, especially sub-micron grains associated with dislocations due to dynamic reshaping of the catalyst particles during the CVD growth of these relatively thick MWCNTs, in addition to interior point defects and surface contamination. These experimental advances lay the foundation for further experimental probing of the phonon-phonon scattering limited thermal conductivity of high-quality quasi-1D SWCNTs.

**Acknowledgements**

The authors thank Karalee Jarvis for advises and helpful discussions of TEM measurements. The experiments are supported by National Science Foundation (NSF) award 2015954 from the Thermal Transport Processes Program. QJ was supported by National Science Foundation (NSF) award 1160494 and a University Continuing Graduate Fellowship of the University of Texas at Austin. The first-principles based calculations are supported by the U.S. Department of Energy, Office of Science, Basic Energy Sciences, Materials Sciences and Engineering Division. The calculations used resources of the Compute and Data Environment for Science (CADES) at the Oak Ridge National Laboratory, which is supported by the Office of Science of the U.S. Department of Energy under Contract No. DE-AC05-00OR22725.

**Reference:**

[1] E. Fermi, P. Pasta, S. Ulam, M. Tsingou, STUDIES OF THE NONLINEAR PROBLEMS, 1955. https://doi.org/10.2172/4376203.

[2] S. Lepri, R. Livi, A. Politi, On the anomalous thermal conductivity of one-dimensional lattices, Europhys. Lett. 43 (1998) 271–276. https://doi.org/10.1209/epl/i1998-00352-3.

[3] O. Narayan, S. Ramaswamy, Anomalous Heat Conduction in One-Dimensional Momentum-Conserving Systems, Phys. Rev. Lett. 89 (2002) 200601. https://doi.org/10.1103/PhysRevLett.89.200601.

[4] J.-S. Wang, B. Li, Intriguing Heat Conduction of a Chain with Transverse Motions, Phys. Rev. Lett. 92 (2004) 074302. https://doi.org/10.1103/PhysRevLett.92.074302.

[5] I. Pomeranchuk, On the Thermal Conductivity of Dielectrics, Phys. Rev. 60 (1941) 820–821. https://doi.org/10.1103/PhysRev.60.820.

[6] L. Lindsay, D.A. Broido, N. Mingo, Lattice thermal conductivity of single-walled carbon nanotubes: Beyond the relaxation time approximation and phonon-phonon scattering selection rules, Phys. Rev. B. 80 (2009) 125407. https://doi.org/10.1103/PhysRevB.80.125407.

[7] N. Mingo, D.A. Broido, Length Dependence of Carbon Nanotube Thermal Conductivity and the "Problem of Long Waves," Nano Lett. 5 (2005) 1221–1225. https://doi.org/10.1021/nl050714d.

[8] Z. Wang, D. Tang, X. Zheng, W. Zhang, Y. Zhu, Length-dependent thermal conductivity of single-wall carbon nanotubes: prediction and measurements, Nanotechnology. 18 (2007) 475714. https://doi.org/10.1088/0957-4484/18/47/475714.

[9] J.A. Thomas, R.M. Iutzi, A.J.H. McGaughey, Thermal conductivity and phonon transport in empty and water-filled carbon nanotubes, Phys. Rev. B. 81 (2010) 045413. https://doi.org/10.1103/PhysRevB.81.045413.

[10] C. Sevik, H. Sevinçli, G. Cuniberti, T. Çağın, Phonon Engineering in Carbon Nanotubes by Controlling Defect Concentration, Nano Lett. 11 (2011) 4971–4977. https://doi.org/10.1021/nl2029333.

[11] K. Sääskilahti, J. Oksanen, S. Volz, J. Tulkki, Frequency-dependent phonon mean free path in carbon nanotubes from nonequilibrium molecular dynamics, Phys. Rev. B. 91 (2015) 115426. https://doi.org/10.1103/PhysRevB.91.115426.

[12] S. Huberman, R.A. Duncan, K. Chen, B. Song, V. Chiloyan, Z. Ding, A.A. Maznev, G. Chen, K.A. Nelson, Observation of second sound in graphite at temperatures above 100 K, Science. 364 (2019) 375–379. https://doi.org/10.1126/science.aav3548.

[13] J. Jeong, X. Li, S. Lee, L. Shi, Y. Wang, Transient Hydrodynamic Lattice Cooling by Picosecond Laser Irradiation of Graphite, Phys. Rev. Lett. 127 (2021) 085901. https://doi.org/10.1103/PhysRevLett.127.085901.

[14] S. Lee, L. Lindsay, Hydrodynamic phonon drift and second sound in a (20,20) single-wall carbon nanotube, Phys. Rev. B. 95 (2017) 184304. https://doi.org/10.1103/PhysRevB.95.184304.

[15] P. Kim, L. Shi, A. Majumdar, P.L. McEuen, Thermal Transport Measurements of Individual Multiwalled Nanotubes, Phys. Rev. Lett. 87 (2001) 215502. https://doi.org/10.1103/PhysRevLett.87.215502.

[16] C. Yu, L. Shi, Z. Yao, D. Li, A. Majumdar, Thermal Conductance and Thermopower of an Individual Single-Wall Carbon Nanotube, Nano Lett. 5 (2005) 1842–1846. https://doi.org/10.1021/nl051044e.

[17] E. Pop, D. Mann, Q. Wang, K. Goodson, H. Dai, Thermal Conductance of an Individual Single-Wall Carbon Nanotube above Room Temperature, Nano Lett. 6 (2006) 96–100. https://doi.org/10.1021/nl052145f.

[18] C.W. Chang, D. Okawa, H. Garcia, A. Majumdar, A. Zettl, Breakdown of Fourier's Law in Nanotube Thermal Conductors, Phys. Rev. Lett. 101 (2008) 075903. https://doi.org/10.1103/PhysRevLett.101.075903.

[19] H. Hayashi, T. Ikuta, T. Nishiyama, K. Takahashi, Enhanced anisotropic heat conduction in multi-walled carbon nanotubes, Journal of Applied Physics. 113 (2013) 014301. https://doi.org/10.1063/1.4772612.

[20] J. Liu, T. Li, Y. Hu, X. Zhang, Benchmark study of the length dependent thermal conductivity of individual suspended, pristine SWCNTs, Nanoscale. 9 (2017) 1496–1501. https://doi.org/10.1039/C6NR06901K.

[21] A.W. Bushmaker, V.V. Deshpande, M.W. Bockrath, S.B. Cronin, Direct Observation of Mode Selective Electron−Phonon Coupling in Suspended Carbon Nanotubes, Nano Lett. 7 (2007) 3618–3622. https://doi.org/10.1021/nl071840f.

[22] I.-K. Hsu, R. Kumar, A. Bushmaker, S.B. Cronin, M.T. Pettes, L. Shi, T. Brintlinger, M.S. Fuhrer, J. Cumings, Optical measurement of thermal transport in suspended carbon nanotubes, Appl. Phys. Lett. 92 (2008) 063119. https://doi.org/10.1063/1.2829864.

[23] I.-K. Hsu, M.T. Pettes, A. Bushmaker, M. Aykol, L. Shi, S.B. Cronin, Optical Absorption and Thermal Transport of Individual Suspended Carbon Nanotube Bundles, Nano Lett. 9 (2009) 590–594. https://doi.org/10.1021/nl802737q.

[24] V.V. Deshpande, S. Hsieh, A.W. Bushmaker, M. Bockrath, S.B. Cronin, Spatially Resolved Temperature Measurements of Electrically Heated Carbon Nanotubes, Phys. Rev. Lett. 102 (2009) 105501. https://doi.org/10.1103/PhysRevLett.102.105501.

[25] M.T. Pettes, L. Shi, Thermal and Structural Characterizations of Individual Single-, Double-, and Multi-Walled Carbon Nanotubes, Adv. Funct. Mater. 19 (2009) 3918–3925. https://doi.org/10.1002/adfm.200900932.

[26] J. Kim, E. Ou, D.P. Sellan, L. Shi, A four-probe thermal transport measurement method for nanostructures, Review of Scientific Instruments. 86 (2015) 044901. https://doi.org/10.1063/1.4916547.

[27] J. Kim, E. Fleming, Y. Zhou, L. Shi, Comparison of four-probe thermal and thermoelectric transport measurements of thin films and nanostructures with microfabricated electro-thermal transducers, J. Phys. D: Appl. Phys. 51 (2018) 103002. https://doi.org/10.1088/1361-6463/aaa9d9.

[28] E. Ou, X. Li, S. Lee, K. Watanabe, T. Taniguchi, L. Shi, Four-Probe Measurement of Thermal Transport in Suspended Few-Layer Graphene With Polymer Residue, Journal of Heat Transfer. 141 (2019) 061601. https://doi.org/10.1115/1.4043167.

[29] B. Smith, G. Fleming, K.D. Parrish, F. Wen, E. Fleming, K. Jarvis, E. Tutuc, A.J.H. McGaughey, L. Shi, Mean Free Path Suppression of Low-Frequency Phonons in SiGe Nanowires, Nano Lett. 20 (2020) 8384–8391. https://doi.org/10.1021/acs.nanolett.0c03590.

[30] E. Fleming, F. Du, E. Ou, L. Dai, L. Shi, Thermal conductivity of carbon nanotubes grown by catalyst-free chemical vapor deposition in nanopores, Carbon. 145 (2019) 195–200. https://doi.org/10.1016/j.carbon.2019.01.023.

[31] B.H. Hong, J.Y. Lee, T. Beetz, Y. Zhu, P. Kim, K.S. Kim, Quasi-Continuous Growth of Ultralong Carbon Nanotube Arrays, J. Am. Chem. Soc. 127 (2005) 15336–15337. https://doi.org/10.1021/ja054454d.

[32] C.L. Cheung, A. Kurtz, H. Park, C.M. Lieber, Diameter-Controlled Synthesis of Carbon Nanotubes, J. Phys. Chem. B. 106 (2002) 2429–2433. https://doi.org/10.1021/jp0142278.

[33] M. Gao, J.M. Zuo, R.D. Twesten, I. Petrov, L.A. Nagahara, R. Zhang, Structure determination of individual single-wall carbon nanotubes by nanoarea electron diffraction, Appl. Phys. Lett. 82 (2003) 2703–2705. https://doi.org/10.1063/1.1569418.

[34] A. Weathers, K. Bi, M.T. Pettes, L. Shi, Reexamination of thermal transport measurements of a low-thermal conductance nanowire with a suspended micro-device, Review of Scientific Instruments. 84 (2013) 084903. https://doi.org/10.1063/1.4816647.

[35] G. Chen, Nanoscale Energy Transport and Conversion: A Parallel Treatment of Electrons, Molecules, Phonons, and Photons., Oxford University Press, 2005.

[36] L. Shi, D. Li, C. Yu, W. Jang, D. Kim, Z. Yao, P. Kim, A. Majumdar, Measuring Thermal and Thermoelectric Properties of One-Dimensional Nanostructures Using a Microfabricated Device, Journal of Heat Transfer. 125 (2003) 881–888. https://doi.org/10.1115/1.1597619.

[37] Y. Touloukian, R. Powell, C. Ho, P. eds Klemens, Thermophysical Properties of Matter, IFI/PLEIMUM • NEW YORK-WASHINGTON, 1970.

[38] L. Lindsay, D.A. Broido, Theory of thermal transport in multilayer hexagonal boron nitride and nanotubes, Phys. Rev. B. 85 (2012) 035436. https://doi.org/10.1103/PhysRevB.85.035436.

[39] J.H. Seol, I. Jo, A.L. Moore, L. Lindsay, Z.H. Aitken, M.T. Pettes, X. Li, Z. Yao, R. Huang, D. Broido, N. Mingo, R.S. Ruoff, L. Shi, Two-Dimensional Phonon Transport in Supported Graphene, Science. 328 (2010) 213. https://doi.org/10.1126/science.1184014.

[40] J. Wang, L. Zhu, J. Chen, B. Li, J.T.L. Thong, Suppressing Thermal Conductivity of Suspended Tri-layer Graphene by Gold Deposition, Adv. Mater. 25 (2013) 6884–6888. https://doi.org/10.1002/adma.201303362.

[41] L. Lindsay, D.A. Broido, N. Mingo, Diameter dependence of carbon nanotube thermal conductivity and extension to the graphene limit, Phys. Rev. B. 82 (2010) 161402. https://doi.org/10.1103/PhysRevB.82.161402.

[42] L. Lindsay, D.A. Broido, N. Mingo, Flexural phonons and thermal transport in multilayer graphene and graphite, Phys. Rev. B. 83 (2011) 235428. https://doi.org/10.1103/PhysRevB.83.235428.

[43] W. Li, J. Carrete, N. A. Katcho, N. Mingo, ShengBTE: A solver of the Boltzmann transport equation for phonons, Computer Physics Communications. 185 (2014) 1747–1758. https://doi.org/10.1016/j.cpc.2014.02.015.

[44] C.D. Landon, N.G. Hadjiconstantinou, Deviational simulation of phonon transport in graphene ribbons with ab initio scattering, Journal of Applied Physics. 116 (2014) 163502. https://doi.org/10.1063/1.4898090.

[45] X. Li, S. Lee, Role of hydrodynamic viscosity on phonon transport in suspended graphene, Phys. Rev. B. 97 (2018) 094309. https://doi.org/10.1103/PhysRevB.97.094309.

[46] X. Li, S. Lee, Crossover of ballistic, hydrodynamic, and diffusive phonon transport in suspended graphene, Phys. Rev. B. 99 (2019) 085202. https://doi.org/10.1103/PhysRevB.99.085202.

[47] X. Li, H. Lee, E. Ou, S. Lee, L. Shi, Reexamination of hydrodynamic phonon transport in thin graphite, Journal of Applied Physics. 131 (2022) 075104. https://doi.org/10.1063/5.0078772.

[48] T.W. Odom, J.L. Huang, P. Kim, C.M. Lieber, Atomic structure and electronic properties of single-walled carbon nanotubes, Nature. 391 (1998) 62–64. https://doi.org/10.1038/34145.

[49] S. Iijima, Helical microtubules of graphitic carbon, Nature. 354 (1991) 56–58. https://doi.org/10.1038/354056a0.

[50] T.W. Ebbesen, P.M. Ajayan, Large-scale synthesis of carbon nanotubes, Nature. 358 (1992) 220–222. https://doi.org/10.1038/358220a0.

[51] S. Tamura, Isotope scattering of dispersive phonons in Ge, Phys. Rev. B. 27 (1983) 858–866. https://doi.org/10.1103/PhysRevB.27.858.

[52] L. Lindsay, W. Li, J. Carrete, N. Mingo, D.A. Broido, T.L. Reinecke, Phonon thermal transport in strained and unstrained graphene from first principles, Phys. Rev. B. 89 (2014) 155426. https://doi.org/10.1103/PhysRevB.89.155426.